\documentclass[twocolumn,showpacs,preprintnumbers,draftclsnofoot]{revtex4}
\usepackage{amsmath}
\usepackage{amssymb}
\usepackage{graphicx}
\usepackage{dcolumn}
\usepackage{bm}
\usepackage{amssymb}
\usepackage{graphicx}
\usepackage{dcolumn}
\usepackage{bm}

\begin{document}

\title{ Topological phase and chiral edge states of bilayer graphene with staggered sublattice potentials and Hubbard interaction }
\author{Ma Luo\footnote{Corresponding author:luom28@mail.sysu.edu.cn} and Zhibing Li}
\affiliation{The State Key Laboratory of Optoelectronic Materials and Technologies \\
School of Physics\\
Sun Yat-Sen University, Guangzhou, 510275, P.R. China}

\begin{abstract}

Gated heterostructures containing bilayer graphene  with  staggered sublattice potentials  are investigated by tight binding model with Rashba spin-orbital coupling and Hubbard interaction. The topological phase diagrams depend on the combinations of substrates and the Hubbard interaction. The presence of the staggered sublattice potential favor the topological phase transition with small Rashba spin-orbital coupling strength. The presence of the Hubbard interaction modified the topological phase boundaries, increasing the minimal spin-orbital coupling strength for topological phase transition. A phase space of topological semi-metal with indirect band gap is identified in the non-interacting systems. For the bilayer graphene with different staggered sublattice potentials in the two layers, the conditions for the zigzag nanoribbons to host edge polarized chiral edge states are discussed. The conditions require moderate or vanishing Rashba spin-orbital coupling strength, as well as proper range of the gate voltage. The conditions for the systems with and without the Hubbard interaction are compared. The edge polarization can be controlled by the gate voltage.

\end{abstract}

\pacs{} \maketitle

\section{Introduction}

Graphene consisted of a single layer or few layers of carbon atoms\cite{neto09,basov14}  is a promising material for advanced electronic\cite{Sarma11}, spintronic and valleytronic devices\cite{weihan14}. The Weyl spinor like carriers near to the Fermi level exhibit high electronic mobility\cite{basov14}, long spin relaxation length and time \cite{HongkiMin06,DanielHuertasHernando06,Boettger07,Tombros07,WeiHan10,Zomer12,TYYang11,WeiHan11,Dlubak12}. Breaking the A-B sublattice symmetry will turn the Weyl spinors to the massive Dirac Fermion like particles. It can be realized by substrates, such as h-BN \cite{Giovannetti07,CRDean10} and SiC\cite{SYZhou07}, which induce staggered sublattice potentials $\Delta$ being equal to 28 meV and 130 meV, respectively. The localized edge states in the zigzag nanoribbons\cite{Eduardo08,Liwei09,Arash16} have been extensively investigated, which are the candidate for information carrier in integrated spintronic and valleytronic devices. The localization of the edge state can be tuned by strain\cite{Stuij15}. We explore the graphene nanoribbons that host edge polarized chiral edge states, whose localization can be controlled by the gated voltage. For this purpose, we firstly study the topological properties of the bulk bilayer graphene with staggered sublattice potentials and Hubbard interaction.


In the AB-stacked bilayer graphene(BLG)\cite{neto09}, an interlayer potential difference $2V$ induced by a gated voltage opens a tunable band gap\cite{Eduardo07}. In the presence of the Rashba spin-orbital coupling(SOC), the gated BLG become topological insulator(TI) with topological invariant $Z_{2}=1$ when the Rashba SOC strength becomes sufficiently large for a given gated voltage \cite{qiao11,qiao13}. Meanwhile, the TI phase has valley Chern number being $C_{V}=2$, implying the quantum spin Hall(QSH) phase. With the Rashba SOC strength smaller than the critical value of topological phase transition, the BLGs is in the quantum valley Hall(QVH) phase with $Z_{2}=0$ and $C_{V}=4$. The Rashba SOC strength induced by external electric fields \cite{HongkiMin06,Yugui07,Zarea09} is far to be sufficient for the TI phase transition. The Rashba SOC could be enhanced by constructing a curve surface \cite{Klinovaja12,Santos16}, substrate proximity effect\cite{Rashba09,Zhenhua10,Jayakumar14,Abdulrhman18} or adatom doping\cite{Fufang07,JunHu12}. In the vicinity of substrate consisted of heavy metal\cite{YuS08,Varykhalov08,Marchenko13}, the crystal field is largely enhanced, which gives rise to sizeable Rashba SOC. The intercalation of heavy metal into the graphene hollow sites can also enhance the Rashba SOC\cite{Marchenko12}. Adding staggered sublattice potentials to both layers of the BLG change the phase diagram, which allow topological phase transition with infinitesimally small Rashba SOC strength at $V\approx\Delta$\cite{Xuechao16}.

In the single-layer graphene(SLG), the presence of SOC induces topologically nontrivial phase, such as QSH phase \cite{CLKane05}. The presence of Rashba SOC modifies the band structure and spin texture, which induces novel optical and electrical properties \cite{XiaojingLi17,DaShuaiMa18}. In the absence of the SOC, the one dimensional armchair nanoribbons exhibit topological band gap as well \cite{TingCao17,Rizzo18}. In comparison, the BLG have more parameters, such as the gated voltage and staggered sublattice potentials of each layer, to control the topological phase. In general, the topological phases of BLG and SLG have different valley Chern numbers.

On the other hand, the electron-electron interaction changes the physical properties of the two dimensional materials, including the topological properties. The Hubbard model has been added to the Kane-Mele (KM) model to incorporate the effect of electron correlation, and it has been  found that the phase diagrams depend on the strength of the Hubbard interaction \cite{StephanRachel10,ShunLi11,Laubach14,Kurita16,Triebl16}. The topological invariant for the interacting systems  can be calculated by integral of the Berry phase that is defined by the eigenstates of the topological Hamiltonian \cite{ZhongWang2010,Gurarie2011,ZhongWang12,ZhongWang122,ZhongWang13,YuanYao16}. The topological Hamiltonian is the inverse of the Green's function at zero frequency. The Green's function of the interacting system can be calculated by cluster perturbation theory(CPT) \cite{Pairault98,Senechal00,Senechal02,Grandi15,Grandi151}.

In this article, the topological phase diagrams of BLGs with different types of staggered sublattice potentials are studied. The effects of electron correlation described by the Hubbard model are calculated by the CPT. The minimal requirement of Rashba SOC strength for topological phase transition is not infinitesimally small, but finite. The effects of next nearest neighbor(NNN) hopping in the tight binding model is also discussed. The conditions for the BLGs that support the edge polarized  chiral edge states within the bulk gap are investigated. The paper is organized as follows. In Sec. II, the tight binding model of the BLGs, the CPT method and the calculation of the topological invariant are described. In Sec. III, the numerical result of the phase diagrams of $Z_{2}$ and $C_{V}$ are discussed. The phase boundaries in the presence and absence of the electron correlation are compared. In Sec. IV, the edge polarized chiral edge states of the zigzag nanoribbons are discussed. In Sec. V, the conclusion is given.

\section{theoretical model}

\subsection{tight binding model}

The AB-stacked bilayer graphene is described by the tight binding Hamiltonian\cite{Tse11,Gelderen10,Konschuh10,Kochan17}
\begin{eqnarray}
H=&&\sum_{l=\pm1}{(H_{l,0}+H_{l,1}+H_{l,R}+H_{l,\Delta_{l}}+H_{l,V}+H_{l,U})} \nonumber
 \\&&+H_{\bot,0}+H_{\bot,1} \label{theHamiltonian}
\end{eqnarray}
where
\begin{equation}
H_{l,0}=-t\sum_{\langle i,j\rangle,\alpha}{(a_{l,i\alpha}^{+}b_{l,j\alpha}+H.c.)}\nonumber
\end{equation}
\begin{equation}
H_{l,1}=-t'\delta_{N}\sum_{\langle\langle i,j\rangle\rangle,\alpha}{(a_{l,i\alpha}^{+}a_{l,j\alpha}+b_{l,i\alpha}^{+}b_{l,j\alpha}+H.c.)}\nonumber
\end{equation}
\begin{equation}
H_{l,R}=ig_{R}\sum_{\langle i,j\rangle,\alpha,\beta}{(a_{l,i\alpha}^{+}[(\mathbf{S}\times\mathbf{d}_{ij})\cdot\hat{z}]_{\alpha\beta}b_{l,j\beta}+H.c.)}\nonumber
\end{equation}
\begin{equation}
H_{l,\Delta_{l}}=\Delta_{l}\sum_{i\in l,\alpha}{(-a_{l,i\alpha}^{+}a_{l,i\alpha}+b_{l,i\alpha}^{+}b_{l,i\alpha})}\nonumber
\end{equation}
\begin{equation}
H_{l,V}=V\sum_{i\in l,\alpha}{l(a_{l,i\alpha}^{+}a_{l,i\alpha}+b_{l,i\alpha}^{+}b_{l,i\alpha})}\nonumber
\end{equation}
\begin{eqnarray}
H_{l,U}=U\sum_{i\in l}[(a_{l,i+}^{+}a_{l,i+}-\frac{1}{2})(a_{l,i-}^{+}a_{l,i-}-\frac{1}{2}) \nonumber \\+(b_{l,i+}^{+}b_{l,i+}-\frac{1}{2})(b_{l,i-}^{+}b_{l,i-}-\frac{1}{2})]\nonumber
\end{eqnarray}
\begin{equation}
H_{\bot,0}=t_{\bot}\sum_{i,\alpha}{(a_{+1,i\alpha}^{+}b_{-1,i\alpha}+H.c.)}\nonumber
\end{equation}
\begin{equation}
H_{\bot,1}=t_{\bot}'\delta_{N}\sum_{\langle\langle i,j\rangle\rangle_{\bot},\alpha}{(b_{+1,i\alpha}^{+}a_{-1,j\alpha}+H.c.)}
\end{equation}
$l=\pm1$ labels two layers, i and j label the primitive cell index, $\alpha=\pm1$ and $\beta=\pm1$ label the spin projections. The operator $a_{l,i\alpha}$($b_{l,i\alpha}$) annihilates a particle at $l$ layer, $i$-th primitive cell and A(B) sublattice with spin $\alpha$. $H_{l,0}$ and $H_{l,1}$ are the nearest neighbor(NN) and NNN hopping Hamiltonians in l layer with $t=2.8$ eV and $t'=0.1$ eV, respectively. The summations with index $\langle i,j\rangle$ and $\langle\langle i,j\rangle\rangle$ run through the NN and NNN sites, respectively. $H_{l,R}$ is the Rashba SOC Hamiltonian with $g_{R}$ being SOC strength, $\mathbf{S}$ being vector of Pauli matrix, $\mathbf{d}_{ij}$ being the unit vector from lattice $i$ to $j$. $H_{l,\Delta_{l}}$ is the staggered sublattice potential for $l$ layer. $\Delta_{+1}$ and $\Delta_{-1}$ take the values from the list [$-130$ meV, $-28$ meV, $0$, $28$ meV, $130$ meV]. $H_{l,V}$ model the potential difference induced by the gated voltage. $H_{l,U}$ is the Hubbard interaction Hamiltonian for half-filling systems with strength being $U$. $H_{\bot,0}$ and $H_{\bot,1}$ are the NN and NNN inter-layer hopping, with $t_{\bot}=0.39$ eV and $t_{\bot}'=0.3$ eV respectively. The summation with index $\langle\langle i,j\rangle\rangle_{\bot}$ runs through the inter-layer NNN sites. The parameter $\delta_{N}$ equates 1 or 0, for the presence or absence of the NNN hopping, respectively. Applying the periodic boundary condition with Bloch phase, the Hamiltonian in the absence of the Hubbard interaction becomes an eight-by-eight matrix. Eigenstates of the matrix equation give the band structure as well as wave functions and Berry phase, which are used to calculate the topological invariant.

\subsection{Cluster perturbation theory}

In the presence of the Hubbard interaction, the single particle eigenstate is not well defined. As a replacement, the single particle Green's function in a unit cell described the dynamic properties of the system. One of the efficient methods to calculate the Green's function with interaction is the CPT. The CPT method has four steps. (1) One  defines an isolated cluster in the lattice. The cluster that usually consists of multiple primitive cells must be able to tile the extended lattice. (2) The Green's function of the cluster with the Hubbard interaction is calculated by exact diagonalization. The Hamiltonian of the isolated cluster is given by Eq. (\ref{theHamiltonian}) with all summations being restricted to the lattice sites within the cluster. Because the Hamiltonian conserves the particle number, the basis states of the many-particle interacting Hamiltonian are the Fock states. For the half-filling N-site cluster, the dimension of the Hilbert space equates to the combination $C^{2N}_{N}$. In this Hilbert space, the Hamiltonian is expressed as a sparse matrix, whose ground state $|\Omega\rangle$ is found by the Lanczos algorithm \cite{Dagotto88}. The cluster Green's function is obtained by the operation
\begin{eqnarray}
G^{C}_{(i,m),(j,n)}(z)&=&\langle\Omega|c_{(i,m)}\frac{1}{z-H+E_0}c_{(j,n)}^{+}|\Omega\rangle \nonumber \\&+&\langle\Omega|c_{(j,n)}^{+}\frac{1}{z+H-E_0}c_{(i,m)}|\Omega\rangle
\end{eqnarray}
where operator $c_{m(n)}$ is the annihilation operator with composite index $m(n)$ for a lattice site within a primitive cell and a spin, $E_0$ is the ground state energy, $z$ is the frequency of the Green's function.  (3) The lattice of  the BLG is covered by isolated clusters, which form a superlattice. The Green's function of the superlattice is given by
\begin{equation}
G^{PC}_{(i,m),(j,n)}(\mathbf{Q},z)=[\frac{G^{C}(z)}{1-V(\mathbf{Q})G^{C}(z)}]_{(i,m),(j,n)}
\label{CPT_superlattice}
\end{equation}
where $\mathbf{Q}$ is the wave vector in the first Brillouin zone of the superlattice, $V(\mathbf{Q})$ is the reciprocal superlattice representation of the hopping matrix between adjacent clusters. Finally, the Green's function of the original lattice is obtained as \cite{Senechal00,Senechal02}
\begin{equation}
G^{CPT}_{m,n}(\mathbf{k},z)=\frac{1}{L}\sum_{i,j=1}^{L}G^{PC}_{(i,m),(j,n)}(\mathbf{k},z)e^{-i\mathbf{k}\cdot(\mathbf{r}_{i}-\mathbf{r}_{j})}
\end{equation}
where $\mathbf{k}$ is the wave vector in the first Brillouin zone of the original lattice, $L$ is the number of primitive cells in the cluster, $\mathbf{r}_{i}$ is the center location of the i-th primitive cell. For the BLGs, the number of lattice sites in a cluster must be  integral multiple  of four, because a primitive cell contains four lattice sites. In our numerical calculation, the isolated cluster contains eight lattice sites(two primitive cells).

\subsection{Topological invariant}

The topological invariant $Z_{2}$ of the band structures is defined as
\begin{equation}
Z_{2}=\frac{1}{2\pi}[\oint_{\partial HBZ}d\mathbf{k}\cdot\mathbf{A}(\mathbf{k})-\int_{HBZ}d^{2}k\Omega_{z}(\mathbf{k})]mod(2)\label{z2int}
\end{equation}
where $HBZ$ means half Brillouin zone, $\mathbf{A}(\mathbf{k})=i\sum_{n}\langle u_{n}(\mathbf{k})|\nabla_{\mathbf{k}}u_{n}(\mathbf{k})\rangle$ is the Berry connection with the summation index n covering all occupied band and $u_{n}(\mathbf{k})$ being the periodic part of the Bloch state, $\Omega_{z}(\mathbf{k})=(\nabla_{\mathbf{k}}\times\mathbf{A})_{z}$ is the z component of the Berry curvature. For the non-interacting case with $U=0$, the Bloch states are given by diagonalization of the Hamiltonian defined in Eq. (\ref{theHamiltonian}) with Bloch periodic condition. For the interacting case with $U\ne0$, the eigenstates of the topological Hamiltonian that equates to inverse of the zero frequency Green's function, $h_{topo}=-[G^{CPT}(\mathbf{k},0)]^{-1}$ \cite{ZhongWang13}, are used in Eq. (\ref{z2int}). Although the eigenstates of $h_{topo}$ are not physical state, they preserve the topological properties of the interacting systems. The numerical integration in Eq. (\ref{z2int}) is performed by Fukui and Hatsugai's procedure \cite{Liang06,Andrew07,Takahiro07,DiXiao10}.

Integrating the Berry curvature through the whole Brillouin zone gives Chern number that is zero, because the Berry curvature is odd under time reversal. The Berry curvature has large magnitude in the vicinity of the K and K' point, so that on can define the continuous Dirac fermion models for each valley. The Berry curvature is defined by the Dirac spinor in momentum space. By integrating the Berry curvatures of the occupied bands through the whole momentum space, $k_{x}\in(-\infty,\infty)$ and $k_{y}\in(-\infty,\infty)$, the Chern number of the continuum Dirac Fermion model corresponding to K and K' valley being denoted as $C_{K}$ and $C_{K'}$, respectively, are defined as
\begin{equation}
C_{K(K')}=\frac{1}{2\pi}\sum_{n=1}^{4}\int_{-\infty}^{\infty}\int_{-\infty}^{\infty}dk_{x}dk_{y}\tilde{\Omega}_{n}^{K(K')}(k_{x},k_{y})
\end{equation}
with the Berry curvatures of the $n$-th band in K(K') valley, $\tilde{\Omega}_{n}^{K(K')}(k_{x},k_{y})$, being given as
\begin{eqnarray}
&&\tilde{\Omega}_{n}^{K(K')}(k_{x},k_{y})= \\
&&-\sum_{n'\ne n}\frac{2Im\langle\psi^{K(K')}_{n\mathbf{k}}|v_{x}|\psi^{K(K')}_{n'\mathbf{k}}\rangle\langle\psi^{K(K')}_{n'\mathbf{k}}|v_{y}|\psi^{K(K')}_{n\mathbf{k}}\rangle}{(\varepsilon_{n'}-\varepsilon_{n})^{2}} \nonumber
\end{eqnarray}
The summation index $n$ covers the four occupied valence bands. $v_{x}$ and $v_{y}$ are the velocity operator of x and y direction in the continuum Dirac Fermion model. $\varepsilon_{n}$ and $|\psi^{K(K')}_{n\mathbf{k}}\rangle$ are the energy level and wave function of the $n$-th band in K(K') valley with wave vector being $\mathbf{k}$. Finally, the valley Chern number is defines as $C_{V}=C_{K}-C_{K'}$ \cite{qiao11,qiao13,Xuechao16,FanZhang13}. The valley Chern number is only calculated in the non-interacting model.

\section{numerical results for the phase diagrams}

In this section, the numerical results for the phase diagrams of the BLGs with different options of substrates are discussed. In the absence of the Hubbard interaction, the phase boundary is determined by the gap closing condition, which can be obtained by solving the non-interacting Hamiltonian with Bloch periodic condition at K point. Two eigenvalues are  $-\Delta_{+1}-V$ and $\Delta_{-1}+V$. The other six eigenvalues are roots of two cubic equations,
\begin{eqnarray}
\Delta_{-1}^{2}\Delta_{+1}+9\Delta_{+1}g_{R}^2+\Delta_{-1}t_{\bot}^2-\Delta_{-1}^{2}V \nonumber \\-9g_{R}^{2}V+t_{\bot}^{2}V-\Delta_{+1}V^{2}+V^{3} \nonumber \\+(-\Delta_{-1}^{2}-9g_{R}^{2}-t_{\bot}^2+2\Delta_{+1}V-V^{2})x \nonumber \\+(-\Delta_{+1}-V)x^2+x^3=0
\end{eqnarray}
and
\begin{eqnarray}
-\Delta_{-1}\Delta_{+1}^{2}-9\Delta_{-1}g_{R}^2-\Delta_{+1}t_{\bot}^2+\Delta_{+1}^{2}V \nonumber \\+9g_{R}^{2}V-t_{\bot}^{2}V+\Delta_{-1}V^{2}-V^{3} \nonumber \\+(-\Delta_{+1}^{2}-9g_{R}^{2}-t_{\bot}^2+2\Delta_{-1}V-V^{2})x \nonumber \\+(\Delta_{-1}+V)x^2+x^3=0
\end{eqnarray}
The gap closing conditions for different combinations of $(\Delta_{+1},\Delta_{-1})$ are obtained by the condition that the two cubic equations have one common root. The NNN hopping terms vanish at K points, so that the presence of the NNN hopping does not change the phase diagram of the non-interacting systems. In the presence of Hubbard interaction, the phase boundary is numerically calculated. The Hubbard interaction coefficient takes the value $U=1.6t$, which have been proven to accurately described the correlation effect in graphene \cite{Wehling11,Schuler13}.

\begin{figure}[tbp]
\scalebox{0.5}{\includegraphics{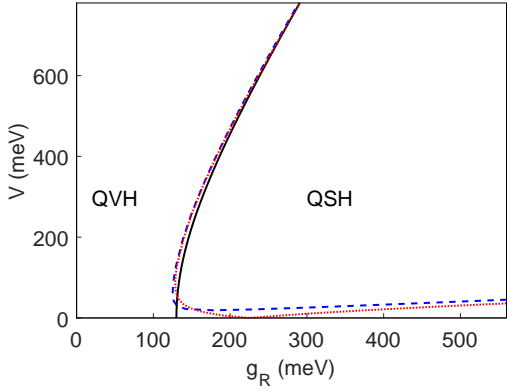}}
\caption{ Phase diagram of suspended BLGs with $\Delta_{+1}=\Delta_{-1}=0$. The phase boundary with black(solid) line is for $(U=0,\delta_{N}=0)$, with blue(dash) line is for $(U=1.6t,\delta_{N}=0)$, with red(dotted) line is for $(U=1.6t,\delta_{N}=1)$.  }
\label{fig_H_Vac}
\end{figure}

The phase diagram of the suspended BLGs is plotted in Fig. \ref{fig_H_Vac}. For the non-interacting systems, the phase boundary between the QSH and QVH phase is $g_{R}=\frac{1}{3}\sqrt{t_{\bot}^{2}+V^{2}}$ \cite{qiao11}. The presence of the Hubbard interaction only modifies the phase boundary  slightly at large gated voltage. The Hubbard interaction expand the regime of QSH at moderate gate voltage($V\approx50\sim300$ meV). The phase boundary is driven above the x axis  by the interaction, which implies that the phase transition to QSH phase requires finite gated voltage for all $g_{R}$. For the interacting system, the presence of the NNN hopping changes the phase boundary for gate voltage being smaller than 50 meV.

\begin{figure}[tbp]
\scalebox{0.39}{\includegraphics{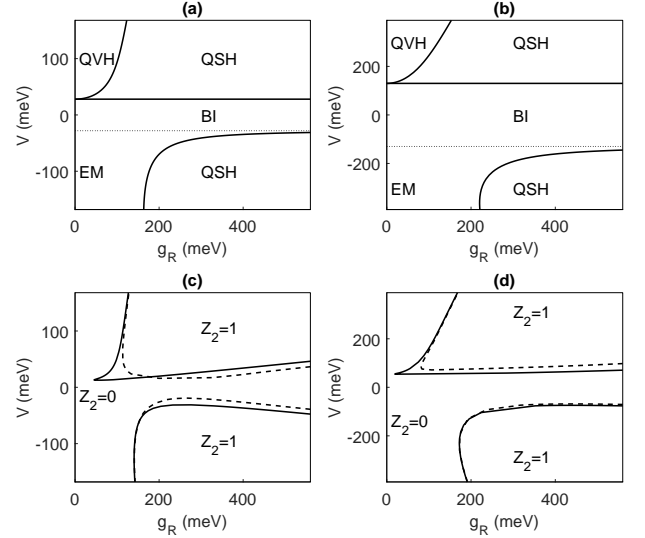}}
\caption{ Phase diagrams of BLGs in the absence of Habbard interaction and the NNN hopping with $\Delta_{+1}=\Delta_{-1}=\Delta=28$ meV in (a), and $\Delta_{+1}=\Delta_{-1}=\Delta=130$ meV in (b). The solid line separate the topological trivial and nontrivial phase with $Z_{2}=0$ and 1, respectively. The dotted line separate the BI and EM phase. The phase diagrams of the same systems with the presence of Habbard interaction and the NNN hopping are plotted in figure (c) and (d). The topological trivial phase regimes are not further divided. The solid and dash lines are the phase boundaries for the systems with $(U=1.6t,\delta_{N}=0)$ and $(U=1.6t,\delta_{N}=1)$, respectively.   }
\label{fig_H_BN}
\end{figure}

The phase diagrams of the BLGs with substrates being $\Delta_{+1}=\Delta_{-1}=\Delta=28$ meV or $\Delta_{+1}=\Delta_{-1}=\Delta=130$ meV in the absent of the Hubbard interaction are plotted in Fig. \ref{fig_H_BN}(a) or (b), respectively. In addition to the QSH and QVH phase, band insulator(BI) phase and edge conductive metal(EM) phase with ($Z_{2}=0$, $C_{V}=0$) appear. The phase boundary between the QSH and QVH phase are \cite{Xuechao16}.
\begin{equation}
g_{R}=\frac{1}{3}\sqrt{\frac{(V-\Delta)(t_{\bot}^{2}+(V+\Delta)^{2})}{V+\Delta}} \label{phaseB1}
\end{equation}
which is plotted as solid lines in Fig. \ref{fig_H_BN}(a) and (b). At $V=\Delta$, the phase boundary approaches the y axis of the phase diagram with $g_{R}=0$, implying topological phase transition at infinitesimally small Rashba SOC strength. However, the presence of the Hubbard interaction modifies the phase boundary, as shown in Fig. \ref{fig_H_BN}(c) and (d), and increases the minimal Rashba SOC strength for topological phase transition to a finite value. Because the valley Chern number is only calculated in the non-interacting model, the phase diagram with interaction only present the phase boundary between topological trivial and nontrivial phase with $Z_{2}=0$ and 1, respectively.  Additional presence of the NNN hopping significantly modifies the phase boundaries for gate voltage with small magnitude.

\begin{figure}[tbp]
\scalebox{0.1}{\includegraphics{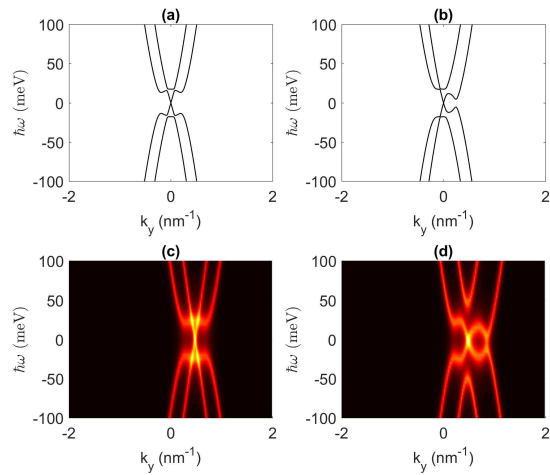}}
\caption{ Band structure of the BLGs with $\Delta_{+1}=\Delta_{-1}=\Delta=130$ meV, $V=148$ meV, $U=0$ in the absence or presence of the NNN hopping in (a) or (b), respectively. The spectral function of the same BLGs with $V=85.5$ meV, $U=1.6t$ in the absence or presence of the NNN hopping in (c) or (d), respectively. The Rashba SOC is tuned to the critical value of the topological phase transition for each system. $k_{y}$ is the wave number along the $K-\Gamma$ line in the first Brillouin zone with the coordinate origin at the K point. $\hbar\omega$ is the energy level of the band structures.  }
\label{fig_spec}
\end{figure}

For the most realistic model that includes both Hubbard interaction and NNN hopping, the phase diagrams show that the minimal Rashba SOC strength along the phase boundaries decrease as the staggered sublattice potentials are increasing. Specifically, for the BLGs with SiC substrates, the minimal Rashba SOC strength along the phase boundary is $g_{R}=81.8$ meV at gate voltage being $V=85.5$ meV. Removing the NNN hopping at the same gate voltage allow the phase transition at smaller Rashba SOC, $g_{R}=60$ meV. For the corresponding non-interacting model, the band structures exhibit linear band crossing at the topological phase transition point, signaling band inversion, as shown in Fig. \ref{fig_spec}(a). The presence of the NNN hopping bring trigonal warping to the band structure, indicated by the local minimal of the first conduction and valence bands beyond the K point, as shown in Fig. \ref{fig_spec}(b). In the presence of the Hubbard interaction and the absence of the NNN hopping, the linear band crossing is clearly visible in Fig. \ref{fig_spec}(c).  The band crossing occur beyond the K point. In the present of the NNN hopping, the two conduction bands and the two valence bands near to the Fermi level strongly couple together, which is indicated by the non-zero spectral function value in the interval among these bands, as shown in Fig. \ref{fig_spec}(d).  The impact of the presence of the interaction is exhibited through the effective Hamiltonian including the self-energy, $H_{eff}=H+\Sigma(\omega,\mathbf{k})$. The self-energy depends on the frequency and Bloch wave vector. The diagonal terms of the self-energy are equivalent to local potentials for the corresponding lattice sites and spin indexes. Among the non-diagonal terms of the self-energy, those corresponding to the first NN hopping terms has the largest magnitude. These terms are equivalent to changing the wave vector in the continuous Dirac Fermion model, so that the band crossing is beyond the K point. The presence of these effective potential and hopping terms change the global properties of the systems. When the gate voltage has large magnitude, the self-energy becomes negligible, so that the phase boundaries of the interacting and non-interacting systems are nearly the same.

\begin{figure}[tbp]
\scalebox{0.39}{\includegraphics{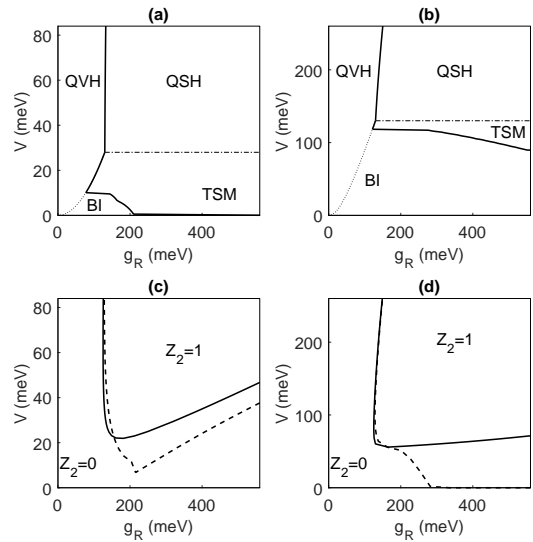}}
\caption{ Same type of plotting as Fig. \ref{fig_H_BN} for the BLGs with $\Delta_{+1}=-\Delta_{-1}=\Delta=28$ meV in (a) and (c), and $\Delta_{+1}=-\Delta_{-1}=\Delta=130$ meV in (b) and (d).   }
\label{fig_H_BNinv}
\end{figure}

Flipping the sign of the staggered sublattice potential of the $l=-1$ layer gives the phase diagrams in Fig. \ref{fig_H_BNinv}. For the non-interacting systems, the phase diagrams are shown in Fig. \ref{fig_H_BNinv}(a) and (b) for BN and SiC substrates, respectively. The topological semi-metal(TSM) phase with $Z_{2}=1$, $C_{V}=0$ and zero indirect band gap appears. Another TSM phase was previously identified in suspended BLGs with both intrinsic and Rashba SOC in reference \cite{HuiPan14}. At the phase boundary between the TSM and BI phases, the gap closing occurs beyond the K point, so that the phase boundary has to be numerically calculated. In the presence of the Hubbard interaction and the NNN hopping, the phase diagrams are shown in Fig. \ref{fig_H_BNinv}(c) and (d). In the regime with $V<\Delta$, the presence of the NNN hopping significantly change the phase boundaries.

The numerical result shows that the BLGs in SiC substrate with $\Delta_{+1}=\Delta_{-1}=\Delta=130$ meV host the topological phase transition with the strength of the Rashba SOC as small as 81.5 meV. The relatively large SOC could be obtain by intercalation of heavy metal element\cite{Marchenko12} between the graphene and SiC substrate or between the two graphene layers. The SOC could be further enhance by applying pressure on the heterostructures that reduce the distance between the graphene and the substrate. An experimental sample of BLG that is encapsulated between two h-BN(SiC) substrates could randomly be BLGs with $\Delta_{+1}=\Delta_{-1}$ or $\Delta_{+1}=-\Delta_{-1}$. The two structures with the same $g_{R}$ have different band gaps, so that they can be distinguished by measuring the band gap of the sample.

For the BLGs with only one substrate, for example, $\Delta_{+1}=\Delta=-28$ meV and $\Delta_{-1}=0$, the phase diagram is similar with that in Fig. \ref{fig_H_BN}(a). For the non-interacting systems, the phase boundary between the QSH and QVH phase are
\begin{eqnarray}
&&g_{R}=\frac{2V-\Delta}{12V} \label{phaseB2}
\\ &&\times\sqrt{\frac{\Delta^{4}+4\Delta^{3}V-16\Delta V^{3}-16V^{2}(t_{\bot}^{2}+V^{2})}{\Delta^{2}-4V^{2}}} \nonumber
\end{eqnarray}
which approaches the y axis of the phase diagram with $g_{R}=0$ at $V=\Delta/2$. The presence of the Hubbard interaction also raises the minimal Rashba SOC strength for the topological phase transition to a finite value. For the BLGs with $|\Delta_{+1}|\ne|\Delta_{-1}|$, the phase diagrams are more complicated. Since these phase diagrams do not contains additional novel phase, the results are not shown in the paper.

The experimental implementations of the BLGs with only one substrate could be obtained by deposition of a suspended BLG on the surface of the SiC or h-BN crystal. Intercalation of heavy metal element in the suspended BLG could be performed before the deposition. After the deposition of the BLG on the substrate, direct growing of another substrate on the top surface of the BLG could be performed. Optionally, the top substrate could also be implemented by a modified scanning tunneling microscope(STM) instrument. Instead of the metallic sharp tip, the tip consisted of SiC or h-BN crystal with atomically flat surface is manufactured. Unlike the regular STM, the insulating tip does not have conducting current for measurement of distance between the tip and the BLG. Instead, the parallelity and distance between the STM flat tip and the BLG  are measured by optical reflection pattern and absorption spectral, respectively. When the STM flat tip and the BLG are parallel, the Fabry-Perot interfere of optical reflection could generate strong signal for spatial adjustment. When the STM flat tip approach the BLG, $\Delta_{+1}$ change from zero to finite value due to the Van der Waals interaction, so that the band gap of the BLG is changed. The longitudinal distance between the STM flat tip and the BLG could be monitored by measuring the optical absorption spectral. Moving the STM flat tip along the transversal direction changed the displacement between the lattices of BLG and STM flat tip, which change $\Delta_{+1}$ between positive and negative values. Thus, the BLG with $\Delta_{+1}=\Delta_{-1}$ and $\Delta_{+1}=-\Delta_{-1}$ could be obtained on demand.

\section{edge polarized chiral edge states of zigzag nanoribbon}

In this section, the edge polarized chiral edge states within the bulk gap is studied. These states exist in the zigzag nanoribbons of BLGs that break the particle-hole symmetric and are in the QVH phase. In the QSH phase, the helical edge states always appear in both two edges. In order to construct the BLGs that support the edge polarized chiral edge states, the non-symmetric staggered sublattice potential($\Delta_{+1}\ne\Delta_{-1}$), finite gate voltage, and moderate or vanishing Rashba SOC are required. We use the BLGs with $\Delta_{+1}=-\Delta_{-1}=\Delta=130$ meV and $g_{R}=28$ meV as example. With the paremters being away from the phase boundaries between QVH and QSH phases, the NNN hopping slightly change the band structure or spectral function, so that it is neglected in the discussion. For the interacting systems, the spectral function of a nanoribbon is obtained by the CPT method. Assuming that the nanoribbon is laid on the x-y plane and the zigzag edges extend along the y axis, the unit cell of the nanoribbon is consisted of N rectangular clusters arranging along the armchair direction along x axis. Each cluster contains eight lattices. The Green's function of the isolated unit cell is obtained by the inversion of a block tri-diagonal matrix with the diagonal block being the inverse of the Green's function of each cluster and the non-diagonal block being the hopping matrix between the neighboring block. The Green's function of the nanoribbon is obtained by Eq. (\ref{CPT_superlattice}).

\begin{figure}[tbp]
\scalebox{0.57}{\includegraphics{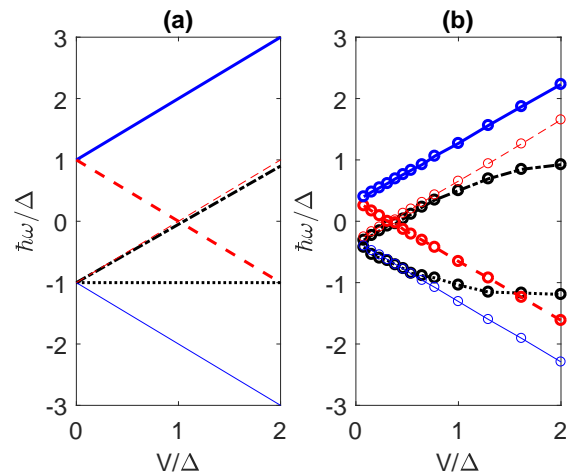}}
\caption{ (a) The energy levels of $F_{L,l}$ and $F_{R,l}$ versus the gate voltage are plotted as solid(blue) and dash(red) lines respectively; The band edges of the conduction and valence bulk band versus the gated voltage are plotted as dash-dot and dotted black lines, respectively, for the BLGs with $\Delta_{+1}=-\Delta_{-1}=\Delta=130$ meV, $g_{R}=28$ meV,$\delta_{N}=0$ and $U=0$. The energy levels of $F_{L,-1}$ and $F_{R,+1}$ are plotted as thin line, because they are always within the valence and conduction bulk band, respectively. (b) The same as (a) with $U=1.6t$. The circle dots are the numerical data, and the lines are guide for eyes.     }
\label{fig_flatB}
\end{figure}

\begin{figure}[tbp]
\scalebox{0.12}{\includegraphics{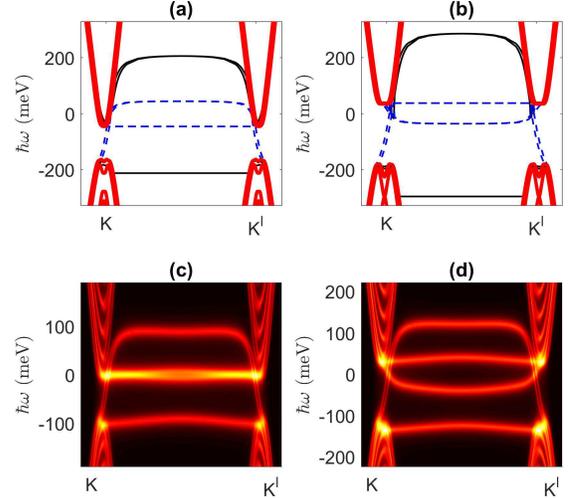}}
\caption{ Band structure of zigzag edge nanoribbon of BLGs with $\Delta_{+1}=-\Delta_{-1}=\Delta=130$ meV, $g_{R}=28$ meV, $U=0$ in (a) $V=84$ meV, and (b) $V=168$ meV. The width of the nanoribbon is 25.56 nm. The bulk band edges is plotted as thick(red) lines. The edge bands localized at the left and right edge are plotted as black(solid) and blue(dash) lines, respectively. Spectral function of the same BLGs with $U=1.6t$ in (c) $V=50$ meV, and (d) $V=84$ meV.  }
\label{fig_edge_3}
\end{figure}

The edge states are originated from the connection between the flat edge bands and the bulk bands. The flat edge bands are localized states at one of the lattice site at the zigzag edge, so that their energy levels are determined by the potential of the corresponding lattice sites. In the absence of the Hubbard interaction, the potentials of the lattice sites are determined by $H_{l,\Delta_{l}}$ and $H_{l,V}$ in the Hamiltonian. In the following discussion, the flat edge band that is localized at the left(right) zigzag edge of the $l$-th layer is denoted as $F_{L,l}$($F_{R,l}$), whose energy level is $\Delta_{l}+lV$($-\Delta_{l}+lV$), as plotted in Fig. \ref{fig_flatB}(a). The band edges of the valence and conduction bulk bands are near to $-\Delta$ and $\Delta+V$.  Meanwhile, $F_{L,l}$($F_{R,l}$) connects to the valence(conduction) bulk bands in the band structures. Assuming positive gate voltage and $\Delta_{+1}=-\Delta_{-1}=\Delta$, the energy level of $F_{L,-1}$($F_{R,+1}$) remains in the valence(conduction) bulk bands, forming none edge state within the bulk gap, as shown by the thin lines in Fig. \ref{fig_flatB}(a). The energy level of $F_{L,+1}$ is within the conduction bulk bands, forming a gapless edge band for arbitrary gated voltage. The weak Rashba SOC split the edge band into two bands. Tuning the gate voltage into the ranges with $\Delta<V$, $\Delta<V<2\Delta$ or $2\Delta<V$ drives the energy level of $F_{R,-1}$ into the conduction bulk band, the bulk gap or the valence bulk band, respectively. Thus, the right zigzag edge hosts none edge states, conductive edge state or gapless edge state, depending on the gate voltage. With $\Delta<V$, the nanoribbon host only edge polarized chiral edge states within the bulk gap, as shown in Fig. \ref{fig_edge_3}(a); with $\Delta<V<2\Delta$, the nanoribbon hosts both of edge polarized  chiral edge states and conductive edge states, as shown in Fig. \ref{fig_edge_3}(b). With $2\Delta<V$, the edge states within the bulk gap is not edge polarized.  Reversing the gate voltage to negative value exchange the properties of $F_{L,l}$ and $F_{R,l}$, so that the localization of the chiral edge states is flipped into the opposite side of the nanoribbon.

In the presence of the Hubbard interaction, the self-energy effectively change the on-site potential of each lattice site, so that the energy levels of the flat edge band and the bulk band are changed. The energy levels of the flat edge bands and the band edges of the conduction and valence bulk band are extracted from the spectral functions of the nanoribbons, which are plotted in Fig. \ref{fig_flatB}(b). The critical values of the gated voltage are obtained from the crossing point between the energy level of $F_{R,-1}$ and the band edge of the bulk bands. With $V\le0.38\Delta$, the energy level of $F_{R,-1}$ remains in the conduction band, so that the nanoribbon host only edge polarized chiral edge states within the bulk gap. The spectral function of the nanoribbon with $V=0.38\Delta$ is plotted in Fig. \ref{fig_edge_3}(c). The flat edge bands become weakly dispersive. With $0.38\Delta<V<1.54\Delta$, the nanoribbon hosts both of edge polarized  chiral edge states and conductive edge states within the bulk gap. The spectral function with $V=0.65\Delta$ is plotted in Fig. \ref{fig_edge_3}(d). Overall, the presence of the Hubbard interaction shrinks the range of the gate voltage that makes the nanoribbon hosting edge polarized chiral edge states, and reduce the effective bulk gap.



\section{conclusion}

The topological phase diagrams of BLG  heterostructures  with optional combinations of h-BN and(or) SiC substrates and the presence of the Hubbard interaction are studied. The Green's function of the interacting systems are calculated by the CPT method. The topological invariants are calculated by employing the topological Hamiltonian that is the inverse of the Green's function at zero frequency. Comparing to the non-interacting systems, the presence of the Hubbard interaction modifies the phase boundaries between topological trivial and non-trivial phases. Specifically, in non-interacting systems, $Z_{2}$ type of topological phase transition can happen at  infinitesimally small Rashba SOC strength. The presence of the Hubbard interaction in the same BLG increases the minimal Rashba SOC strength for topological phase transition to a finite value. For BLGs with $\Delta_{+1}=-\Delta_{-1}$ in the absence of the Hubbard interaction, topological semi-metal(TSM) phase with zero indirect band gap and non-trivial topological invariants, i.e. $Z_{2}=1$ and $C_{V}=0$, is founded.

The conditions that the zigzag nanoribbon of the BLGs can host edge polarized chiral edge states are studied. The staggered sublattice potentials of the two layers are required to be asymmetric. For the typically choice with $\Delta_{+1}=-\Delta_{-1}=\Delta$, in the absence of the Hubbard interaction, the BLGs with $0<V<\Delta$ host pure edge polarized chiral edge states, and the BLGs with $\Delta<V<2\Delta$ host the mixing of edge polarized chiral edge states and conductive edge states. In the presence of the Hubbard interaction, the range of the gate voltage for these two phases shrink. The localization of the edge polarized chiral edge states can be controlled by the sign of the gated voltage. These features would provide more feasible systems for the design of spintronic and valleytronic devices.

\begin{acknowledgments}
The project is supported by the National Natural Science Foundation of China (Grant No.
11704419), the National Basic Research
Program of China (Grant No. 2013CB933601), and the National Key Research and Development Project of China
(Grant No. 2016YFA0202001).
\end{acknowledgments}

\clearpage

\end{document}